\documentclass[aps,showkeys,showpacs,twocolumn,floatfix]{revtex4}
\usepackage{graphicx}
\usepackage{amsmath}
\usepackage{amsfonts}
\usepackage{color}

\begin{document}

\title{Majority-vote model on $(3,12^2)$, $(4,6,12)$ and $(4,8^2)$ Archimedean lattices}

\author{\bf F. W. S. Lima}
\email{fwslima@gmail.com}
\affiliation{
Dietrich Stauffer Computational Physics Lab, Departamento de F\'{\i}sica,
Universidade Federal do Piau\'{\i},\\
64.049-550, Teresina, Piau\'{\i}, Brazil
}

\date{\today}

\begin{abstract}
On ($3,12^2$), ($4,6,12$) and ($4,8^2$) Archimedean lattices, the critical properties of majority-vote model are considered and studied  using the Glauber transition rate proposed by Kwak {\it et all.} [Phys. Rev. E, {\bf 75}, 061110 (2007)] rather than the traditional majority-vote with noise [Jos\'e M\'ario de Oliveira, J.
Stat. Phys. {\bf 66}, 273 (1992)]. The critical temperature and the critical exponents for this 
transition rate are obtained from extensive Monte Carlo simulations and with a finite size scaling analysis.
The calculated values of the critical temperatures Binder cumulant are $T_c=0.363(2)$ and $U_4^*=0.577(4)$; $T_c=0.651(3)$ and $U_4^*=0.612(5)$; and $T_c=0.667(2)$ and $U_4^*=0.613(5)$ for ($3,12^2$), ($4,6,12$) and ($4,8^2$) lattices, respectively.
The critical exponents $\beta/\nu$, $\gamma/\nu$ and $1/\nu$ for this model are 
$0.237(6)$, $0.73(10)$, and $ 0.83(5)$; $0.105(8)$, $1.28(11)$, and $1.16(5)$; 
$0.113(2)$,
$1.60(4)$, and $0.84(6)$ for ($3,12^2$), ($4,6,12$) and ($4,8^2$) lattices, respectively.
These results differ from the usual Ising model results and the majority-vote model on so-far studied regular lattices or complex networks.
\end{abstract}

\pacs{
 05.10.Ln, %% Monte Carlo methods statistical physics and nonlinear dynamics
 05.70.Fh, %% Phase transitions
 64.60.Fr  %% Critical exponents
}

\keywords{Monte Carlo simulation, critical exponents, phase transition, non-equilibrium}

\maketitle

%% ############################################################################
\section{Introduction}
%% ############################################################################

The use of local majority rules to study voting systems was introduced by Galam three decades ago to study bottom-up democratic voting in hierarchical structures \cite{sg1}. It is one of the founding papers of sociophysics with a follow-up paper published a few years latter in the Journal of Statistical Physics \cite{sg2}, which 33 years later has devoted a special issue to the modelling of social systems \cite{sg3}  including a paper by Galam extending his earlier work from two to three parties \cite{sg4}. Indeed, while sociophysics has been rejected by physicists in the eighties \cite{sg5}, it is has become today an active field of research among physicists all over the world \cite{sg3, ds, sg6}. 

The local majority rule model has motivated a good deal of works under several names including the Majority Model and the Majority Vote Model (MVM).
The nonequilibrium majority-vote model proposed by Oliveira \cite{MVM-SL} defined on two-dimensional regular lattices shows second-order phase transition with critical exponents $\beta$, $\gamma$, $\nu$ identical \cite{MVM-SL,MVM-regular,MVM-MFA} with those of equilibrium Ising model \cite{ising,critical} that agree with hypothesis of Grinstein {\it et al.} \cite{grinstein}.

The MVM on the complex networks exhibit different behavior \cite{MVM-SW0,MVM-SW1,MVM-ERU,MVM-ERD,MVM-VD,MVM-ABU,MVM-ABD, MVM-APN}.
Campos {\it et al.} investigated MVM on {\it undirected} small-world network \cite{MVM-SW0}.
They found that the critical exponents $\gamma/\nu$ and $\beta/\nu$ are different from those of the Ising model \cite{critical} and depend on the rewiring probability.
Luz and Lima studied MVM on {\it directed} small-world network \cite{MVM-SW1} constructed using the same process described by S\'anchez {\it et al.} \cite{sanchez}.
They found that the critical exponents $\gamma/\nu$ and $\beta/\nu$ also are different from those of the Ising model on square lattices and in this case MVM the exponents do not depend on the rewiring probability, that is contrary to results of Campos {\it et al.} \cite{MVM-SW0} .
Pereira {\it et al.} \cite{MVM-ERU} and Lima {\it et al.} \cite{MVM-ERD} studied MVM on {\it undirected} Erd{\H o}s--R\'enyi's (ERU) on {\it directed} Erd{\H o}s--R\'enyi's (ERD) classical random graphs \cite{ER} and their results obtained for critical exponents agree with the results of Pereira {\it et al.} \cite{MVM-ERU}, within the error bars. After
Lima {\it et al.} \cite{MVM-VD} also studied the MVM on random Voronoy--Delaunay lattice \cite{VD} with periodic boundary conditions.
Lima also \cite{MVM-ABD,MVM-ABU} studied the MVM on {\it directed} Albert--Barab\'asi (ABD) and {\it undirected} Albert--Barab\'asi (ABU)  network \cite{AB} and contrary to the Ising model on these networks \cite{alex}, the order/disorder phase transition {\em was} observed in this system.
However, the calculated $\beta/\nu$ and $\gamma/\nu$ exponents for MVM on ABD and ABU networks are different from those for the Ising model \cite{critical} and depend on the mean value of connectivity $\bar z$ of ABD and ABU network.
Lima and Malarz \cite{lima-malarz} studied the MVM on $(3,4,6,4)$ and $(3^4,6)$ Archimedean lattices (AL).
They remark that the critical exponents $\gamma/\nu$, $\beta/\nu$ and $1/\nu$ for MVM on $(3,4,6,4)$ AL are {\em different} from the Ising model \cite{critical} and {\em differ} from those for so-far studied regular two-dimensional lattices \cite{MVM-SL,MVM-regular}, but for $(3^4,6)$ AL, the critical exponents are much closer to those known analytically for SL Ising model. Santos {\it et all.} \cite{santos} studied the MVM on triangular ($3^{6}$), honeycomb ($6^{3}$) and
Kagom\'e ($3,6,3,6$) AL. They found for $(3^6)$, $(6^3)$ and $(3,6,3,6)$ AL some critical exponents are much closer to those known analytically for square lattice Ising model, i.e. $\beta=1/8=0.125$, $\gamma=7/4=1.75$ and $\nu=1$, but except for $\nu$ they differ for more than three numerically estimated uncertainties.
\begin{figure*}[!hbt]
\bigskip
\begin{center}
\includegraphics[angle=0.7, width=0.70\textwidth]{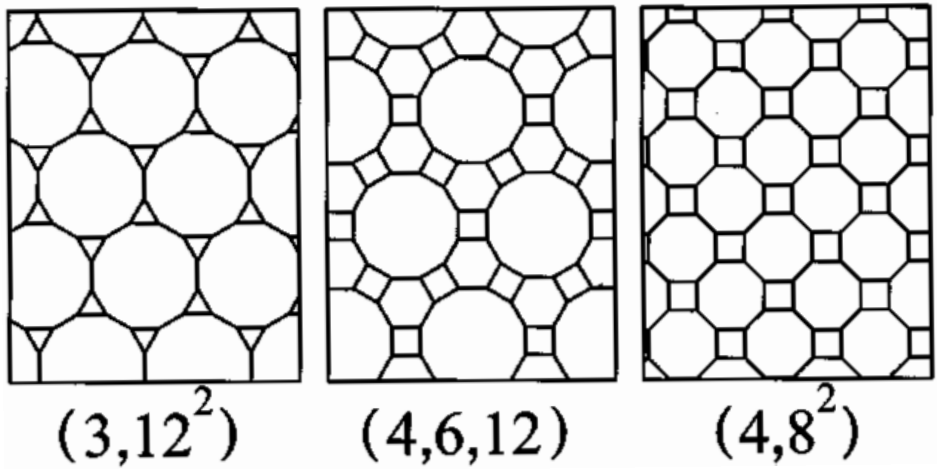}
\end{center}
\caption{\label{fig-AL} Display the picture of the $(3,12^2)$, $(4,6,12)$ and $(4,8^2)$ AL, respectively.}
\end{figure*}
%% ----------------------------------------------------------------------------
\bigskip

The results presented in Refs. \cite{MVM-SW0,MVM-SW1,MVM-ERU,MVM-ERD,MVM-VD,MVM-ABU,MVM-ABD, MVM-APN} show that the MVM on various complex topologies belongs to different universality classes.
Moreover, contrary for MVM on regular lattices \cite{MVM-SL,MVM-regular}, the obtained critical exponents are different from those of the equilibrium Ising model \cite{critical}.
Very recently, Yang and Kim \cite{yang2008} showed that also for $d$-dimensional hypercube lattices ($3\le d\le 6$) critical exponents for MVM differ from those for SL Ising model.
The same situation occurs on hyperbolic lattices \cite{wu2010}.

In this paper we study the MVM on three AL, namely $(3,12^2)$, $(4,6,12)$, and $(4,8^2)$.
The AL are vertex transitive graphs that can be embedded in a plane such that every face is a regular polygon.
The AL are labeled according to the sizes of faces incident to a given vertex.
The face sizes are sorted, starting from the face for which the list is the smallest in lexicographical order.
In this way, the lattice gets the name $(3,12^2)$, $(4,6,12)$ and $(4,8^2)$.
Critical properties of these lattices were investigated in terms of site percolation \cite{percolation} and Ising model \cite{zborek}.

Our main goal is to check the hypothesis of Grinstein {\it et al.} \cite{grinstein}, i.e., that non-equilibrium stochastic spin systems with up-down symmetry fall into the universality class of the equilibrium Ising model on regular lattices (like SL \cite{MVM-SL}) and complex spin systems (like spins on ERU and ERD \cite{MVM-ERU,MVM-ERD} or ABU and ABD \cite{MVM-ABU,MVM-ABD}).

With extensive Monte Carlo simulation we show that MVM on $(3,12^2)$, $(4,6,12)$ and $(4,8^2)$ AL exhibits second-order phase transitions and has critical exponents that {\em do not} fall into the universality class of the equilibrium Ising model. The picture the $(3,12^2)$, $(4,6,12)$ and $(4,8^2)$ AL are showed in the 
Fig. \ref{fig-AL}.
%% ----------------------------------------------------------------------------
\begin{figure*}[!hbt]
\begin{center}
\includegraphics[width=0.9\textwidth]{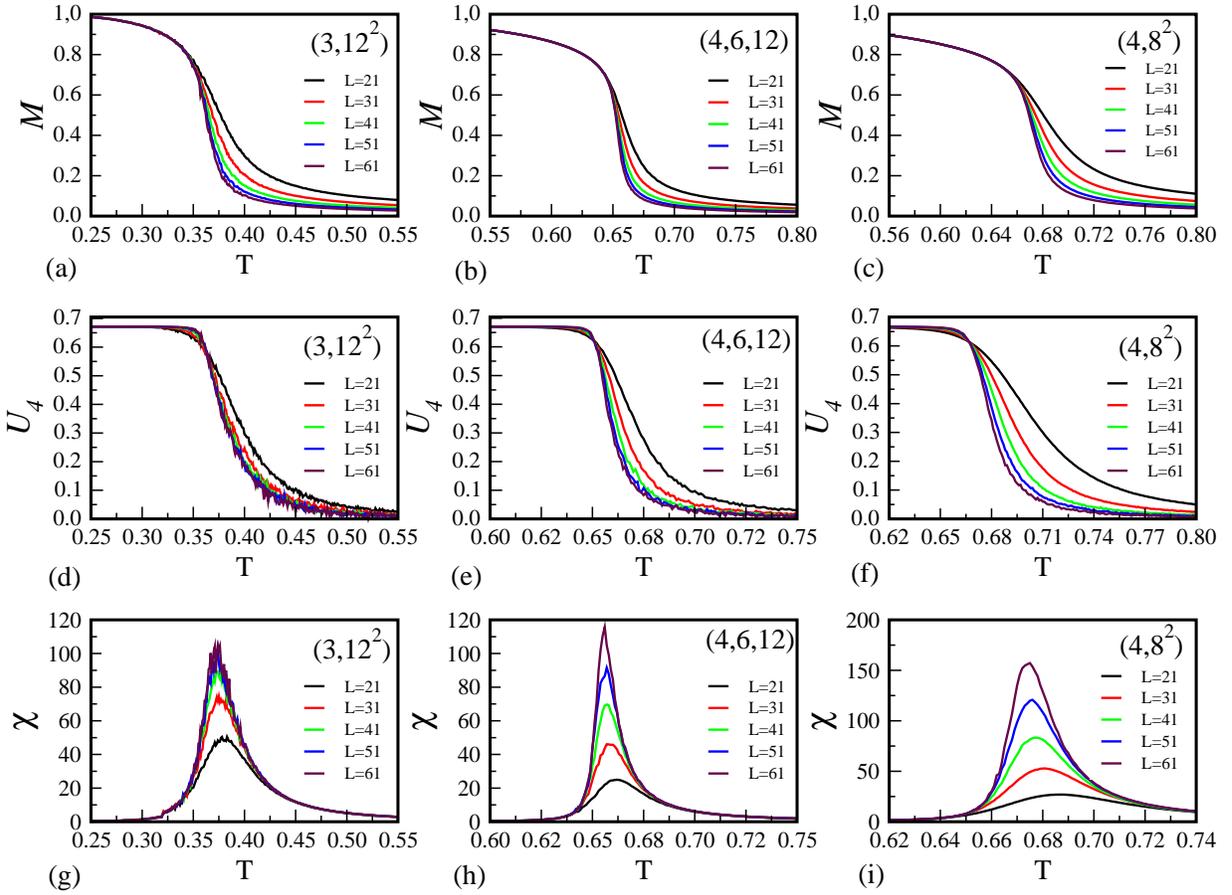}
\end{center}
\caption{\label{fig-M-U-C} (Color on-line). The magnetization $M$, Binder cumulant $U_4$, and susceptibility $\chi$ as a function of the temperature $T$, for $L=21$, $31$, $41$, $51$, and $61$ lattice sizes, and with $N=6L^2$ sites for $(3,12^2)$ (first column), $N=12L^2$ $(4,6,12)$ (second column) and $N=4L^2$ sites for $(4,8^2)$ AL (third column).}
\end{figure*}
%% ----------------------------------------------------------------------------
%% ############################################################################
\section{Model and simulation}
%% ############################################################################

On the original MVM \cite{MVM-SL}, the system dynamics is as follows.
Initially, we assign a spin variable $\sigma$ with values $\pm 1$ at each site
of the square lattice (SL). At each step we try to flip the spin
of the nodes in a sequential way. The flip is accepted with
probability
\begin{equation} 
w_i=\frac{1}{2}\left[ 1-(1-2q)\sigma_{i}\cdot\text{S}\left(\sum_{j=1}^k\sigma_j\right)\right],
\end{equation}
where $S(x)$ is the sign $\pm 1$ of $x$ if $x \neq 0$ and $S(x) = 0$ if $x = 0$.
To calculate $w_{i}$ our sum runs over the number $k$ ($k=4$ for SL) of nearest
neighbors of {\it i}th spin. Equation (1) means that with probability
$(1 - q)$ the spin will adopt the same state as the majority of its
neighbors. Here, the control parameter $0\leq q \leq 1$ plays a role
similar to that of the temperature in equilibrium systems. The
smaller the $q$ the greater the probability of parallel aligning
with the local majority \cite{MVM-SL,MVM-ERU,MVM-ERD,MVM-VD,MVM-ABD,MVM-ABU}.

Here we study the MVM on $(3,12^2)$, $(4,6,12)$ and $(4,8^2)$ AL using a alternative probability
of Eq. (1) called Glauber rate
probability proposed by Kwak {\it et all.} \cite{Kwak}. The Glauber transition rates of MVM 
can be written as
\begin{equation} 
w_{GL}=\frac{1}{2}\left[ 1- \sigma_{i}\cdot\text{S}\left(\sum_{j=1}^k\sigma_j\right)tanh\beta_{T}\right],
\end{equation}
where $\beta_{T}$ is the inverse of the temperature $1/K_{B}T$ and $K_{B}$ is the Boltzmann constant.
Comparing this expression with Eq. (1), we see the correspondence between the original MVM and
that with Glauber dynamics, which leads to the relation between the noise parameter $q$ and the temperature in Glauber dynamics as $(1-2q)=tanh\beta_{T}$.

To study the critical behavior of the model we define the variable $m\equiv\sum_{i=1}^{N}\sigma_{i}/N$.
In particular, we are interested in the magnetization $M$, susceptibility $\chi$ and the reduced fourth-order cumulant $U$
\begin{subequations}
\label{eq-def}
\begin{equation}
M(T)\equiv \langle|m|\rangle,
\end{equation}
\begin{equation}
\chi(T)\equiv N\left(\langle m^2\rangle-\langle m \rangle^2\right),
\end{equation}
\begin{equation}
U(T)\equiv 1-\dfrac{\langle m^{4}\rangle}{3\langle m^2 \rangle^2},
\end{equation}
\end{subequations}
where $\langle\cdots\rangle$ stands for a thermodynamics average.
The results are averaged over the $N_{\text{run}}$ independent simulations.

These quantities are functions of temperature $T$ and obey the finite-size scaling relations
\begin{subequations}
\label{eq-scal}
\begin{equation}
\label{eq-scal-M}
M=L^{-\beta/\nu}f_m(x),
\end{equation}
\begin{equation}
\label{eq-scal-chi}
\chi=L^{\gamma/\nu}f_\chi(x),
\end{equation}
\begin{equation}
\label{eq-scal-dUdq}
\frac{dU}{dT}=L^{1/\nu}f_U(x),
\end{equation}
where $\nu$, $\beta$, and $\gamma$ are the usual critical 
exponents, $f_{m,\chi,U}(x)$ are the finite size scaling functions with
\begin{equation}
\label{eq-scal-x}
x=(T-T_c)L^{1/\nu}
\end{equation}
\end{subequations}
being the scaling variable. Therefore, from the size dependence of $M$ and $\chi$ we obtained the exponents $\beta/\nu$ and $\gamma/\nu$, respectively. The maximum value of susceptibility also scales as $L^{\gamma/\nu}$. Moreover, the value of $T^*$ for which $\chi$ has a maximum is expected to scale with the lattice size as
\begin{equation}
\label{eq-q-max}
T^*=T_c+bL^{-1/\nu} \text{ with } b\approx 1.
\end{equation}
Therefore, the relations \eqref{eq-scal-dUdq} and \eqref{eq-q-max} may be used to get the exponent $1/\nu$.

We performed Monte Carlo simulation on the $(3,12^2)$, $(4,6,12)$ and $(4,8^2)$ AL with various lattice of size ($L=21$, $31$, $41$, $51$, and $61$) for $(3,12^2)$ with $N=6xL^{2}$ that give $N=2646$, $5766$, $10086$, $15606$ and $ 22306$; $(4,6,12)$ with $N=12xL^{2}$ and $N=5292$, $11532$, $20172$, $31212$, and $44652$; and for $(4,8^2)$ with $N=4xL^2$ and $1764$, $3844$, $6724$, $10404$, and $14884$ sites.
It takes $2\times 10^5$ Monte Carlo steps (MCS) to make the system reach the steady state, and then the time averages are estimated over the next $2\times 10^5$ MCS.
One MCS is accomplished after all the $N$ spins are investigated whether they flip or not.
The results are averaged over $N_{\text{run}}$  $(30\le N_{\text{run}} \le 50)$ independent simulation runs for each lattice and for given set of parameters $(T,N)$.
%% ----------------------------------------------------------------------------
%% ----------------------------------------------------------------------------
\begin{table}
\caption{\label{tab} Critical parameter, exponents and effective dimension for MVM model on $(3,12^2)$, $(4,6,12)$ and $(4,8^2)$ AL.
For completeness we cite data for SL Ising model as well.}
%% scale-free AB and random ER graphs with an average connectivity $\bar z=4$ and also for $(3,4,6,4)$ and $(3^4,6)$ AL.
\begin{ruledtabular}
\begin{tabular}{r lllllll}
%%	& SL \cite{MVM-SL} 
%%	& AB \cite{MVM-ABD} 
%%	& ER \cite{MVM-ERU} 
%%	& (3,4,6,4)\cite{lima-malarz} 
%%	& $(3^4,6)$ \cite{lima-malarz} 	
	& $(3,12^2)$ 
	& $(4,6,12)$ 
	& $(4,8^2)$
	& SL Ising \\
\hline
$T_c$
%%      & 0.075(10)
%%	& 0.431(3) 
%%	& 0.181(1) 
%%	& 0.091(2)  
%%	& 0.134(3) 
	& 0.363(2) 
	& 0.651(3) 
	& 0.667(2)
	& ~ \\
$\beta/\nu$
%%	& 0.125(5)  
%%	& 0.447(1) 
%%	& 0.242(6) 
%%	& 0.103(6)  
%%	& 0.114(3) 
	& 0.237(6) 
	& 0.105(8) 
	& 0.113(2)
	& 0.125 \\
$\gamma/\nu$\footnote{obtained using $\chi(N)$ at $T=T_c$}     
%%	& 1.73      
%%	& 0.104(2) 
%%	& 0.54(1)  
%%	& 1.60(5) 
%%	& 1.64(3) 
	& 0.73(10) 
	& 1.28(11) 
	& 1.60(4) 
	& 1.75 \\
$\gamma/\nu$\footnote{obtained using $\chi(N)$ at $T=T^*$}
%%	& 1.70     
%%	& 0.89(9) 
%%	& 0.52(6) 
%%	& 1.66(5) 
%% 	& 1.68(3) 
	& 0.70(8) 
	& 1.44(4) 
	& 1.66(2)
	& 1.75 \\
$1/\nu$
%%	& 1.01(5) 
%%	& ---      
%%	& 0.59(7)  
%%	& 0.88(8) 
%%	& 0.98(1) 
	& 0.83(5) 
	& 1.16(5) 
	& 0.84(6)
	& 1 \\
\end{tabular}
\end{ruledtabular}
\end{table}
%% ----------------------------------------------------------------------------
%% ############################################################################
\section{Results and Discussion}
%% ############################################################################

%% ----------------------------------------------------------------------------

In Fig. \ref{fig-M-U-C} we show the dependence of the magnetization $M$, Binder cumulant $U_4$, and the susceptibility $\chi$ on the temperature $T$, obtained from simulations on $(3,12^2)$, $(4,6,12)$ and $(4,8^2)$  AL with $N$ ranging from $N=1764$ to $44652$ sites.
The shape of $M(T)$, $U(T)$, and $\chi(T)$ curve, for a given value of $N$, suggests the presence of the second-order phase transition in the system.
The phase transition occurs at the value of the critical temperature $T_c$.
The critical noise parameter $T_c$ is estimated as the point where the curves for different system sizes $N$ intercept each other \cite{binder}.
Then, we obtain $T_c=0.363(2)$ and $U_4^*=0.577(4)$; $T_c=0.651(3)$ and $U_4^*=0.612(5)$; $T_c=0.667(2)$ and $U_4^*=0.613(8)$ for $(3,12^2)$, $(4,6,12)$ and $(4,8^2)$ AL, respectively.

\begin{figure}[!hbt]
\bigskip
\includegraphics[width=0.4\textwidth]{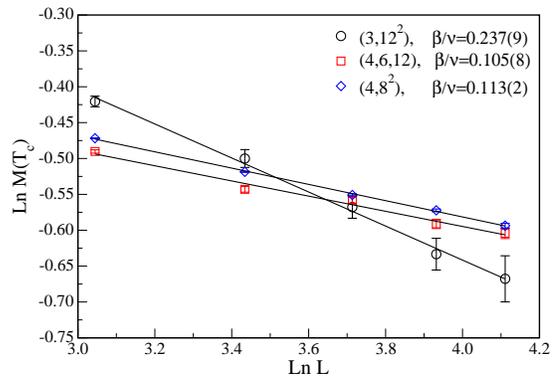}
\caption{\label{fig-M} Plot of the magnetization $M^*=M(T_c)$ vs. the linear system size $L$.}
\end{figure}

In Fig. \ref{fig-M} we plot the dependence of the magnetization $M^*=M(T_c)$ vs. the linear system size $L$.
The slopes of curves correspond to the exponent ratio $\beta/\nu$ according to Eq. \eqref{eq-scal-M}.
The obtained exponents are $\beta/\nu=0.237(6)$, $0.105(8)$, and $0.113(2)$, respectively for $(3,12^2)$, $(4,6,12)$ and $(4,8^2)$ AL.
%% ----------------------------------------------------------------------------
%% ----------------------------------------------------------------------------
%% ----------------------------------------------------------------------------
\begin{figure}[!hbt]
\bigskip
\includegraphics[width=0.4\textwidth]{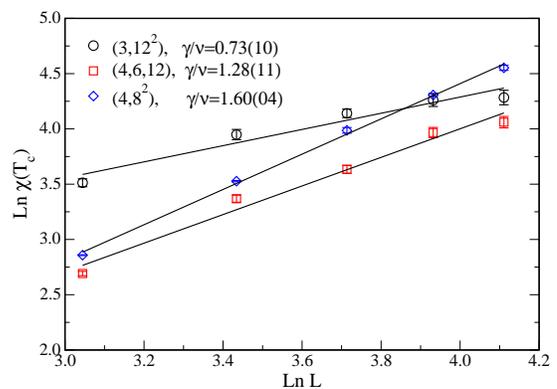}
\caption{\label{fig-chi-N1} Susceptibility at $T_c$ versus $L$ for  $(3,12^2)$, $(4,6,12)$ and $(4,8^2)$ AL.}
\end{figure}
%% ----------------------------------------------------------------------------
\begin{figure}[!hbt]
\bigskip
\includegraphics[width=0.40\textwidth]{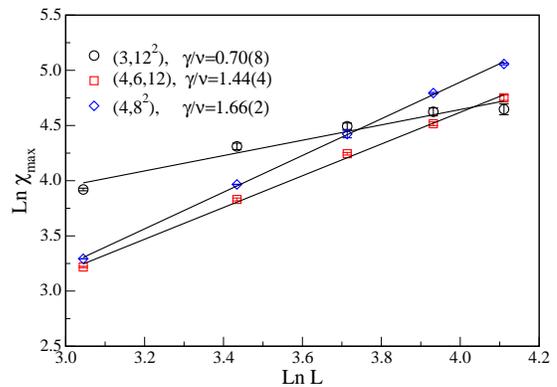}
\caption{\label{fig-chi-N2} Susceptibility at $T_{\chi_{max}}(N)$ versus $L$ for $(3,12^2)$, $(4,6,12)$ and $(4,8^2)$ AL.}
\end{figure}
\begin{figure}[!hbt]
\includegraphics[width=0.40\textwidth]{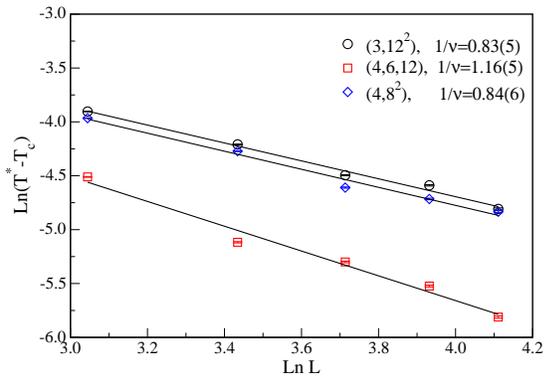}
\bigskip
\caption{\label{fig-eq-4} Plot $\ln|T_c(L)-T_c|$ versus the linear system size $L$ for $(3,12^2)$ (circles), $(4,6,12)$ (squares), $(4,8^2)$ (diamonds).}
\bigskip
\end{figure}
\begin{figure}[!hb]
\includegraphics[width=0.40\textwidth]{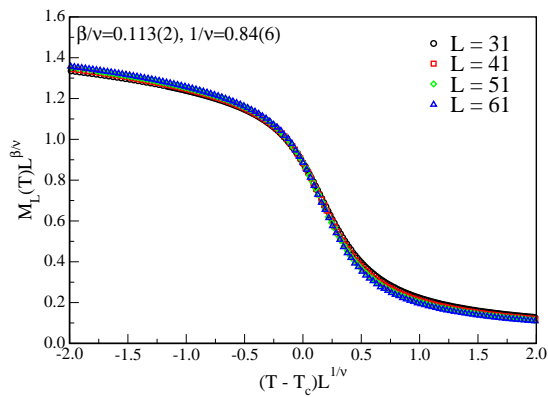}
\caption{\label{fig-eq-cl1} Data colapse of the magneisation {\it M} shown in Fig. \ref{fig-M-U-C}(c) for the linear system size $L=31,41,51$, and $61$ for $(4,8^2)$ AL. The exponents used here were
$\beta/\nu=0.113(2)$ and $1/\nu=0.84(6)$ }
\end{figure}
\bigskip
\begin{figure}[!hb]
\includegraphics[width=0.40\textwidth]{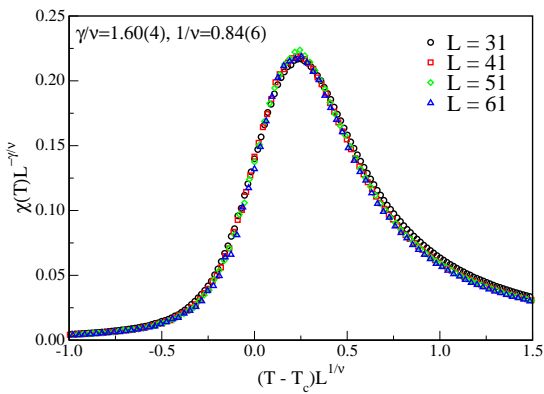}
\bigskip
\caption{\label{fig-eq-cl2} Data colapse of the susceptibility shown in Fig. \ref{fig-chi-N1}(i) for the linear system size $L=31,41,51$, and $61$ for $(4,8^2)$ AL. The exponents used here were
$\gamma/\nu=1.66(2)$ and $1/\nu=0.84(6)$.}
\end{figure}

The exponents ratio $\gamma/\nu$ at $T_{c}$ are obtained from the slopes of the straight lines with $\gamma/\nu=0.73(10)$ for $(3,12^2)$, $\gamma/\nu=1.28(11)$ for $(4,6,12)$, and $\gamma/\nu=1.60(4)$ for $(4,8^2)$, as presented in Fig. \ref{fig-chi-N1}. The exponents ratio $\gamma/\nu$ at $T_{\chi_{max}}(N)$ are $\gamma/\nu=0.70(8)$ for $(3,12^2)$, $\gamma/\nu=1.44(4)$ for $(4,6,12)$, and $\gamma/\nu=1.66(2)$ for $(4,8^2)$, as presented in Fig. \ref{fig-chi-N2}.

To obtain the critical exponent $1/\nu$, we used the scaling relation \eqref{eq-q-max}.
The calculated  values of the exponents $1/\nu$ are $1/\nu=0.83(5)$ for $(3,12^2)$ (circles),
 $1/\nu= 1.16(5)$ for $(4,6,12)$ (squares), and $1/\nu=0.84(6)$ for $(4,8^2)$ (diamonds) (see Fig. \ref{fig-eq-4}).
%% ----------------------------------------------------------------------------
%% ---------------------------------------------------------------------------- 

We plot $ML^{\beta/\nu}$ versus $(T-T_{c})L^{1/\nu})$ in the Fig. \ref{fig-eq-cl1} using the critical
exponents $1/\nu=0.113(2)$ and $\beta/\nu=0.84(6)$ for size lattice $L=31,41,51$, and $61$ for 
for $(4,8^2)$ AL. The excellent collapse of the curves for four different system sizes corroborates the extimation for $T_c$ and the critical exponents $\beta/\nu$ and $1/\nu$.
%% ----------------------------------------------------------------------------
In the Fig. \ref{fig-eq-cl2} we plot $\chi L^{-\gamma/\nu}$ versus $(T-T_{c})L^{1/\nu})$ using the critical exponents $\gamma/\nu=1.60(4)$ and $\beta/\nu=0.84(6)$ for size lattice $L=31,41,51$, and $61$ 
for $(4,8^2)$ AL. Again, the excellent of the curves for four different system sizes corroborates the extimation for $T_c$ and the critical exponents $\gamma/\nu$ and $1/\nu$.The results of simulations are collected in Tab. \ref{tab}.

%% ############################################################################
\section{Conclusion}
%% ###########################################################################

Finally, we remark that the critical exponents $\gamma/\nu$, $\beta/\nu$ and $1/\nu$ for MVM on {\em regular} $(3,12^2)$, $(4,6,12)$ and $(4,8^2)$ AL are {\em similar} to the MVM model on {\em regular} $(6^3)$, $(3,6,3,6)$ and $(3^6)$ \cite{santos} and also at $(3,4,6,4)$ and $(3^4,6)$ \cite{lima-malarz} and are {\em different} from the Ising model \cite{critical} and {\em differ} from those for so-far studied regular lattices \cite{MVM-SL,MVM-regular} and for the {\it directed} and {\it undirected} ER random graphs \cite{MVM-ERU,MVM-ERD} and for the {\it directed} and {\it undirected} AB networks \cite{MVM-ABD,MVM-ABU}. 
However, in the latter cases \cite{MVM-ERU,MVM-ERD,MVM-ABU,MVM-ABD} the scaling relations \eqref{eq-scal} must involve the number of sites $N$ instead of linear system size $L$ as these networks in natural way do not posses such characteristic which allow for $N\propto L^d$ $(d\in\mathbb{Z})$ dependence \footnote{The linear dimension of such networks, i.e. its diameter --- defined as an average node-to-node distance --- grows usually logarithmically with the system size \cite{el-sw}.}.
For $(3,12^2)$, $(4,6,12)$ and $(4,8^2)$ AL some critical exponents are different to those known analytically for square lattice Ising model, i.e. $\beta=1/8=0.125$, $\gamma=7/4=1.75$ and $\nu=1$.
%% ############################################################################
\begin{acknowledgments}
F.W.S.L. is grateful to Dietrich Stauffer for stimulating discussions and for critical reading of the manuscript.
F.W.S.L. acknowledge also the support the system SGI Altix 1350 the computational park CENAPAD, UNICAMP-USP, SP-BRASIL and also the agency FAPEPI for the financial support.
\end{acknowledgments}
 
%% ############################################################################

\end{document}